\documentclass{ro}
\usepackage{pstricks}
\usepackage{fancybox}
\usepackage{amssymb}
\usepackage{amsmath}
\usepackage{graphicx}
\usepackage[utf8]{inputenc}
\usepackage{epsfig,psfrag,graphics,multirow}

\def \afig#1#2 {\begin{figure}[!hp]
\begin{center} 
\mbox{\psfig{file=#1.eps}}
\end{center}
\caption{#2}
\label{fig:#1}
\end{figure}}

\def\bbbn{\rm I\!N}
\newenvironment{bulletitemize}{%

\begin{itemize}}{\end{itemize}}

\newtheorem{remark}{Remark}

\newtheorem{thm}{Theorem}
\newtheorem{lem}{Lemma}
\newtheorem{cor}{Corollary}

\newenvironment{pf}{%

\textit{Proof}}
{
\hfill$\square$
}

\newcommand {\subsetsum}{\textsc{subset sum~}}
\newcommand {\multiplesubsetsum}{\textsc{multiple subset sum~}}
\newcommand {\multiplesubsetsumdc}{\textsc{multiple subset sum with different knapsack capacities~}}
\newcommand {\multipleknapsackassignmentrestrictions}{\textsc{multiple knapsack assignment restriction~}}
\newcommand {\ssp}{\textsc{ss}}

\newcommand{\MSS}{\textsc{mss}}
\newcommand{\MSSDC}{\textsc{mssdc}}
\newcommand{\MKAR}{\textsc{mkar}}
\newcommand {\SAToneinthree}{\textsc{one-in-($2$,$3$)sat($2$,$\bar{1}$)}}
\newcommand{\nphard}{$\mathcal{NP}$-hard }

\graphicspath{{figures/}}

\begin{document}

\title{Some complexity and approximation results for coupled-tasks scheduling problem according to topology}

\author{B. Darties}\address{LE2I, UMR CNRS 6306, University of Burgundy,\\8 Rue Alain Savary 21000 Dijon, France}
\author{R. Giroudeau}\address{LIRMM, UMR CNRS 5506, Universit\'e de Montpellier\\161 rue Ada, 34392 Montpellier Cedex 5, France}
\author{J.-C. K\"{o}nig}\address{LIRMM, UMR CNRS 5506, Universit\'e de Montpellier\\161 rue Ada, 34392 Montpellier Cedex 5, France}
\author{G. Simonin}\address{Insight Centre for Data Analytics, University College Cork, Ireland}


\begin{abstract}
We consider the  makespan minimization  coupled-tasks problem  in presence of compatibility constraints with a specified topology. In particular, we focus on stretched coupled-tasks, {\it  i.e.} coupled-tasks having the same sub-tasks execution time and idle time duration. We study several problems in framework of classic complexity and approximation for which the compatibility graph is bipartite (star, chain, $\ldots$). In such a context, we design some efficient polynomial-time approximation algorithms  for an intractable scheduling problem according to some parameters.
\end{abstract}

\maketitle 



\section{Introduction and model}

The detection of an object by a common radar system is based on the following principle: a transmitter emits a uni-directional pulse that propagates though the environmental medium. If the pulse encounters an object, it is reflected back to the transmitter. Using the transmit time and the  direction of the pulse, the position of the object can be computed by the transmitter. Formally this acquisition process is divided into three parts: (i) pulse transmission, (ii) wave propagation and reflection, (iii) echo reception. Thus the detection system must perform two tasks (parts (i) and (iii)) separated by an idle time (part (ii)). Such systems generally run in non-preemptive mode: once started, a task cannot be interrupted  and resumed later. However, the idle time of an acquisition task can be reused to perform another task. On non-preemptive mono-processor systems, scheduling issues appear when in parallel several sensors using different frequencies are working: the idle time of an acquisition task can be reused to perform partially on entirely a second acquisition process using another sensor, but only if both sensors use different frequencies to avoid interferences. Otherwise these two acquisitions processes should be scheduled sequentially.

Coupled-tasks, introduced first by Shapiro \cite{Shapiro}, seem to be a natural way to model, among others, data acquisition processes: a coupled-task $t_i$ is composed by two sub-tasks with processing time $a_i$ and $b_i$ and whose execution must be separated by an incompressible and not flexible time $l_i$ (called the idle time of the task). For an acquisition process, a sensor emits a radio pulse as a first sub-task, and listens for an echo reply as a second sub-task, while the radio pulse propagation operates during an idle time $l_i$.

Coupled-tasks are also an efficient way to model acquisition systems designed to detect changes in an environment for a given period, by producing two measurements before and after the given period. Here each measurement can be modeled as a sub-task.


%
%

We note  $\mathcal{T}=\{t_1,\ldots,t_n\}$ the collection of coupled-tasks to be scheduled. In order to minimize the makespan (schedule length) of $\mathcal{T}$, it is necessary to execute one or several different sub-tasks during the idle time of a coupled-task. In the original model, all coupled-tasks may be executed in each other according to processing time of sub-tasks and the duration of the idle time.

Some papers investigated the problem of minimizing the makespan for various configurations depending on the values of $a_i$, $b_i$ and $l_i$ \cite{ABGOR04,BEKPTW09,op1997}. In \cite{op1997}, authors present a global visualization of scheduling problems complexity with coupled-tasks, and give main complexity results.

In a multi-sensors acquisition system, incompatibilities may arise between two tasks $t_i$ and $t_j$ if they operate with two different sensors working at the same channel. Thus any valid schedule would require $t_i$ and $t_j$ to be scheduled sequentially. Hereafter, we propose a generalization of an original coupled-tasks model by considering the notion of  compatibility constraint among tasks: original coupled-task model, by introducing compatibility constraint among tasks:
two tasks $t_i$ and $t_j$ are \textit{compatibles} if any sub-task of $t_i$ may be executed during  the idle time of $t_j$ or reciprocally. In \cite{sdgk10journal}, we introduced a compatibility graph $G=(V, E)$ to model such this compatibility, where $V=\mathcal{T}$ is the entire collection of coupled-tasks, and each pair of compatible tasks are linked by an edge $e\in E$. We proposed in \cite{sdgk10journal,sdgk11journal} new results focused on the impact of the addition of $G$ on the complexity of the problem. 


Our work is motivated by the acquisition of data for automatic vehicle under water, as a TAIPAN torpedo. With the growth in robotic technologies, several applications and works are emerging and the theoretical needs are a priority. For example, the torpedo is used to execute several submarine topographic surveys, including topological or temperature measurements. These acquisitions tasks can be partitioned into specific sub-problems, where their modelling is very precise. 

Since the engineers have a wide degree of freedom to create and transform the different tasks, they required a strong theoretical analysis of coupled tasks with compatibility constraint. Indeed, they needed to have a better knowledge of the difficulty of scheduling coupled-tasks on such systems, and to compare their scheduling heuristics to the optimal one.


\subsection{Contribution}
\label{notations}
In this work, we propose new results of complexity and approximation for  particular problem instances composed by \textit{stretched coupled-tasks} only : a stretched coupled-task is a coupled-task $t_i=(a_i, l_i, b_i)$ for which the three parameters $a_i$, $b_i$ and $l_i$ are equal to the same value $\alpha(t_i)$, called the \textit{stretch factor} of $t_i$.  

We investigate here the problem of scheduling on a mono-processor a set of stretched coupled-tasks, subject to compatibility constraint in order to minimize the completion time of the latest task. For clarity, $a_i$ (resp $b_i$) refers either to the first (resp. second) sub-task, or to its processing time according to the context.

A major research issue concerns the impact of the class of the compatibility graph $G$ on the complexity of the problem: it is known that the problem is \nphard even when all the tasks are compatibles between each other, \textit{i.e.}  $G$ is a complete graph (see \cite{op1997}). On the other side, when $G$ is an empty graph a trivial optimal solution would consist in scheduling tasks sequentially. Our aim is to  determine the complexity of the problem when $G$ describes some sub-classes of bipartite graphs, and to propose approximation algorithms with performance guarantee for \nphard instances.

\begin{remark}
\label{remark:3alpha}
If two compatibles stretched coupled-tasks $t_i$ and $t_j$, with $\alpha(t_i) \leq \alpha(t_j)$, are scheduled in parallel in any solution of the scheduling problem, then one of the following conditions must hold: 
\begin{enumerate}
\item either $\alpha(t_i) = \alpha(t_j)$: then the idle time of one task is fully exploited to schedule a sub-task from the other (i.e. $b_i$ is scheduled during $l_j$, and $a_j$ is scheduled during $l_i$), and the execution of the two tasks is done without idle time.
\item or $3\alpha(t_i) \leq \alpha(t_j)$: then $t_i$ is fully executed during the idle time $l_j$ of $t_j$. For sake of simplify, we say we \textit{pack} $t_i$ into $t_j$. 
\end{enumerate}
The others configuration $\alpha(t_i) < \alpha(t_j) < 3\alpha(t_i)$ is unavailable, otherwise some sub-tasks would overlap in the schedule.
\end{remark}

From Remark \ref{remark:3alpha} one can propose an orientation to each edge $e=(t_i, t_j)\in E$ from the task with the lowest stretch factor to the task with the highest one, or set $e$ as a bidirectional edge when $\alpha(t_i)=\alpha(t_j)$. In the following, we consider only oriented compatibility graphs. Abusing notation, dealing with undirected topologies for $G$ refers in fact to its underlying undirected graph.

We use various standard notations from graph theory: $N_{G}(x)$ is the set of neighbors of $x$ in $G$. $\Delta_{G}$ is the maximum degree of $G$.
We denote respectively by $d^-_{G}(v)$ and  $d^+_{G}(v)$ the indegree and outdegree of $v$, and $d_{G}(v)=d^-_{G}(v)+d^+_{G}(v)$.  We denote by $G[S]$ the graph induced from $G$ by vertices from $S$.

Reusing the Graham's notation scheme \cite{GLLRK79}, we define the main problem of this study as $1\vert\alpha(t_i),G \vert C_{max}$. We study the variation of the complexity when the topology of $G$ varies, and we propose approximation results for \nphard instances.

We study thee subclasses of bipartite graphs in particular: the chain, the star, and the k-stage bipartite graphs.
A $k$-stage bipartite graph is a graph $G=(V_0 \cup V_1 \cup\dots \cup V_k, E_1 \cup E_2\cup \ldots \cup E_k)$, where  each arc in $E_i$ has its extremities in $V_{i}$ and in $V_{i+1}$, for $i \in \{1,\ldots,k\}$. 
For a given k-stage bipartite graph $G$, we denote by $G_k=G[V_{k-1} \cup V_k]$ the $k$th stage of $G$. 
In this paper, we focus our study on $1$-stage bipartite graphs ($\boldsymbol{1}$\textbf{-SBG}) and $2$-stage bipartite graphs ($\boldsymbol{2}$\textbf{-SBG}). 
We also study the problem when the compatibility graph $G$ is a 1-stage complete bipartite graph ($\boldsymbol{1}$\textbf{-SCBG}), i.e. $E_1$ contains all the edges $(x,y)$, $\forall x\in V_0$, $\forall y\in V_1$. 

For $1$-SBG (or $2$-SBG) with $G=(X\cup Y,E)$, we denoted by $X$-tasks (resp. $Y$-tasks) the set of tasks represented by $X$ (resp. $Y$) in $G$.
For any set of $X$-tasks, let $seq(X)$ be the time required to schedule sequentially all the tasks from $X$. Formally, we have:
$$seq(X)=\sum_{t\in X} 3\alpha(t).$$

\begin{remark}
\label{rem:IS}
Given an instance of $1\vert\alpha(t_i),G \vert C_{max}$. If $X$ is an independent set for $G$, then all the tasks from $X$ are pairwise non-compatibles. Thus $seq(X)$ is a lower bound for the cost of any optimal solution.
\end{remark}

The results obtained in this article are summarized in Table \ref{tab:recap}.

\begin{table}[!h]
\centering
       \begin{tabular}{|p{4cm}|p{4cm}|p{4cm}|}
       \hline     
\textbf{Topology}& \textbf{Complexity} &\textbf{Approximation}\\
\hline
$G$=Chain graph & $O(n^3)$ (Theo. \ref{chaintheo})&\\
  \hline
\hline 
$G$=Star graph \footnotemark[1]& $\mathcal{NP-C}$ (Theo. \ref{startheo}) & $\mathcal{FPTAS}$ (Theo. \ref{startheoapprox})\\ 
\hline
$G$=Star graph \footnotemark[2]& $O(n)$ (Theo. \ref{corostar}) & \\ 
  \hline
  \hline
$G$= $1$-SBG, $d_G(Y)\leq 2$ & $O(n^3)$ (Theo.  \ref{polybipartite}) & \\
  \hline
 \multirow{2}{*}{$G$= $1$-SCBG} & \multirow{2}{*}{$\mathcal{NP-C}$ (see \cite{sgk10})}& $\mathcal{PTAS}$ (Theo. \ref{alltheorems})\\
&  & $\frac{13}{12}$-$\mathcal{APX}$(Theo. \ref{alltheorems})\\
\hline
$G$= $2$-SBG  & $\mathcal{NP-C}$ (Theo. \ref{bipartitheo}) & $\frac{13}{9}$-$\mathcal{APX}$   (Theo. \ref{th:2stage:apx13/9})\\
  \hline
\end{tabular}
\caption{Complexity and approximation results discussed in this paper.} 
\label{tab:recap}
\end{table}

\footnotetext[1]{Star graph with only incoming arcs for the central node arc.}   
\footnotetext[2]{Star graph with at least one outcoming arc for the central node.}

\subsection{Prerequisites}
\label{prerequisites}

\subsubsection{Performance ratio}

Recall that the performance ratio $\rho$ for a minimization (resp. maximization) problem is given as the ratio between the value of the approximation solution returned by the algorithm $\mathcal{A}$ on an instance $\mathcal{I}$ and the optimum \textit{i.e.} $\rho  \leq \max_{\mathcal{I}} \frac{\mathcal{A(I)}}{\mathcal{OPT(I)}}$ (resp. $\rho \geq \min_{\mathcal{I}}  \frac{\mathcal{OPT(I)}}{\mathcal{A(I)}}$). Notice that for a minimization problem the ratio is greater than one (resp. lower than one).

\subsubsection{Definition of problems}
To prove the different results announced in this paper, we use several well-known approximation results on four packing-related problems: 
\begin{enumerate}
\item The \subsetsum (\ssp) problem is a well-known problem in which, given a set $\mathcal{S}$ of $n$ positive values and $v \in \bbbn$, one asks if there exists  a subset $\mathcal{S^*}\subseteq \mathcal{S}$ such that $\sum_{i\in \mathcal{S^*}} i = v$. 
This decision problem is well-known to be ${\mathcal{NP}}$-complete (see \cite{GJ79}). The optimization version problem is sometimes viewed as a \textsc{knapsack} problem, where each item profits and weights coincide to a value in $\mathcal{S}$, the knapsack capacity is $v$, and the aim is to find  the set of packable items with maximum profit.

\item The \multiplesubsetsum  (\MSS) problem is a variant of well-known \textsc{bin packing} in which a number of identical bins is given and one would like to maximize the overall weight of the items packed in the bins  such that the sum of the item weights in every bin does not exceed the bin capacity. 
The problem is also a special case of the \textsc{Multiple knapsack} problem in which all knapsacks have the same capacities and the item profits and weights coincide. Caprara et al. \cite{CKP00} proved that \MSS ~admits a $\mathcal{PTAS}$, but does not admit a $\mathcal{FPTAS}$ even for only two knapsacks. They also proposed a $\frac{3}{4}-$approximation algorithm in \cite{CKP03}.

\item \multiplesubsetsumdc (\MSSDC) ~\cite{CKP002} is an extension of \MSS~considering different bin capacities. \MSSDC ~also admits a $\mathcal{PTAS}$ \cite{CKP002}. 
\item As a generalization of \MSSDC, \multipleknapsackassignmentrestrictions (\MKAR) problem consists to pack weighted items into non-identical capacity-constrained bins, with the additional constraint that each item can be packed in some bins only. Each item as a profit, the objective here is to maximize the sum of profits of packed items. 
Considering that the profit of each item equals its weight, \cite{DKKSR00} proposed a $\frac{1}{2}$-approximation.
\end{enumerate}

We also use a well-known result concerning a variant of the ${\mathcal{NP}}$-complete problem \textsc{3SAT} \cite{GJ79}, denoted subsequently  by \SAToneinthree: 
An instance of \SAToneinthree ~is described by the following elements: we use  ${\mathcal{V}}$ to denote the set of $n$ variables. Let $n$ be a multiple of $3$ and let ${\mathcal{C}}$ be a set of clauses of cardinality $2$ or $3$. There are $n$ clauses of cardinality $2$ and $n/3$ clauses of cardinality $3$ such that:
\begin{itemize}
\item Each clause of cardinality $2$ is equal to $(x \vee \bar{y})$ for some $x,~y \in {\mathcal{V}}$ with $x \neq y$.

\item Each of the $n$ literals $x$ (resp. of the $n$ literals $\bar{x}$) for $x \in {\mathcal{V}}$ belongs to one of the $n$ clauses of cardinality $2$, thus to only one of them.
\item Each of the $n$ (positive) literals $x$ belongs to one of the $n/3$ clauses of cardinality $3$, thus to only one of them.

\item Whenever $(x \vee \bar{y})$ is a clause of cardinality $2$ for some $x,~y \in {\mathcal{V}}$, then $x$ and $y$ belong to different clauses of cardinality $3$.
\end{itemize}
 The aim of \SAToneinthree ~is to find if there exists a truth assignment $I:{\mathcal{V}} \rightarrow \{0,1\}$, $0$ for false and $1$ for true, whereby each clause in ${\mathcal{C}}$ has exactly one true literal. \SAToneinthree ~has been proven ${\mathcal{NP}}$-complete in \cite{gkmptcs08}.

As an example, the following logic formula is the smallest valid instance of \SAToneinthree :
$(x_0 \vee x_1 \vee x_2) 
\wedge (x_3 \vee x_4 \vee x_5) \wedge (\bar{x}_0 \vee x_3) \wedge (\bar{x}_3 \vee x_0) \wedge (\bar{x}_4 \vee x_2) 
\wedge (\bar{x}_1 \vee x_4) \wedge (\bar{x}_5 \vee x_1) \wedge (\bar{x}_2 \vee x_5)$. 

The answer to \SAToneinthree ~is $yes$. It is sufficient to choose $x_0=1$, $x_3=1$ and $x_i=0$ for $i=\{1,2,4,5\}$. This yields a truth assignment that satisfies the formula, and  there is exactly one true literal for each  clause.

\section{Computational complexity for some classes of compatibility graphs}
\label{complexitesection}

In this section, we present two preliminary results of complexity for the problem that consists in scheduling a set of stretched-coupled tasks with compatibility constraints. In such a context, we will consider the topologies of chain and star.

First we show that the problem is solvable within a $O(n^3)$ time complexity algorithm when $G$ is a chain (Theorem \ref{chaintheo}). 
Then we prove that it is \nphard even when the compatibility graph is a star (Theorem \ref{startheo}), 

\subsection{Chain graph}

Despite of the simplicity of a chain topology, solving  the scheduling problem on a chain is not as simple as it appears : a main issue arise when two adjacent vertices $x$ and $y$ have the same stretch factor. In this configuration, we cannot determine locally if $x$ and $y$ can be packed together in an optimal solution or not (this requires to examine the neighbourhood of $x$ and $y$, and this problematic configuration can be repeated all along the chain).  However, we show that the scheduling problem with a chain is polynomial using a similar method as developed in \cite{sdgk11journal}.  

\begin{thm}
\label{chaintheo}
The problem  $1| \alpha(t_i), G=chain | C_{max}$ admits a polynomial-time algorithm.
\end{thm}

\begin{pf}
This problem can be solved in polynomial-time by a reduction to the search for a minimum weighted perfect matching. This problem can be polynomially solved in $O(n^3)$ time complexity \cite{Edmonds}. 

First, note that if for a task $x$ with two neighbors $y$ and $z$, we have  $3(\alpha(y) + \alpha(z)) \leq  \alpha(x)$, the idle duration of $x$ is high enough to schedule both $y$ and $z$. Thus one can schedule $y$ and $z$ into $x$ without decreasing the cost of any optimal solution, and remove tasks $x$, $y$ and $z$ from the studied graph. Thus, in the rest of the proof, one can restrict our study to chains  $G=(V,E)$ such that for any $x\in V$, we have  $3\sum_{y\in N_G(x)}\alpha(y) >  \alpha(x)$.

In order to obtain a graph with an even number of vertices and to find a perfect matching, we construct a graph $H=(V_H,E_H,w)$ and we define a weighted function $w:E\rightarrow \bbbn$ as follows: 
\begin{enumerate}
\item Let $\mathcal{I}_1$ be an instance of our problem with a compatibility graph $G=(V,E)$, and $\mathcal{I}_2$ an instance of the minimum weight perfect matching problem in graph constructed from $\mathcal{I}_1$. We consider a graph $H$, consisting of two copies of $G$ denoted by $G'=(V',E')$ and $G''=(V'',E'')$. The vertex corresponding to $x\in V$ is denoted by $x'$ in $G'$ and $x''$ in $G''$. Moreover, $\forall i =1,\ldots, n$, an edge $\{x', x''\}$ in $E_H$ is added and we state 
$w(\{x', x''\})=\boldsymbol{3\times \alpha(x)'}$. This weight represents the sequential time of the $x'$-task.
We have $H=G'\cup G''=(V'\cup V'',E'\cup E'')$, with $\vert V'\cup V''\vert$ of even size.

\item For two compatible tasks $x'$ and $y'$ with $3\times \alpha_{x'} \leq \alpha_{y'} $ or $3\times \alpha_{y'} \leq \alpha_{x'}$, 
we add the edges  $\{x', y'\}$ and  $\{x'', y''\}$ in $E_H$ and we state $w(\{x', y'\})=w(\{x'', y''\})=\boldsymbol{\frac{3\times \max \{ \alpha_{x'}, \alpha_{y'} \}}{2}}$.

\item For two compatible tasks $x'$ and $y'$ with $\alpha_{x'}\! =\! \alpha_{y'} $, we add the edges $\{x', y'\}$ and  $\{x'', y''\}$ in
 $E_H$, and we state $w(\{x', y'\})=w(\{x'', y''\})=\boldsymbol{\frac{4\times \alpha_{x'}}{2}}$.
\end{enumerate}

Figure~\ref{fig:ExGcc2} shows an example of construction of $H$ when $G$ is a chain with $3$ vertices.

\psfrag{A1'}{\scriptsize{$x_1'$}}
\psfrag{A2'}{\scriptsize{$x_2'$}}
\psfrag{A3'}{\scriptsize{$x_3'$}}
\psfrag{A1''}{\scriptsize{$x_1''$}}
\psfrag{A2''}{\scriptsize{$x_2''$}}
\psfrag{A3''}{\scriptsize{$x_3''$}}

\psfrag{2p+2p}{\scriptsize{$3\times 2$}}
\psfrag{2p+3p}{\scriptsize{$3\times 8$}}
\psfrag{2p+p-3}{\scriptsize{$3\times 8$}}
\psfrag{Gc}{\scriptsize{$G$}}
\psfrag{Gc'}{\scriptsize{$G'$}}
\psfrag{Gc''}{\scriptsize{$G''$}}
\psfrag{Hc}{\scriptsize{$H$}}

\psfrag{2p}{\scriptsize{$\alpha_1=2$}}
\psfrag{3p}{\scriptsize{$\alpha_3=8$}}
\psfrag{p-3}{\scriptsize{$\alpha_2=8$}}
\psfrag{val1}{\scriptsize{$\frac{3 \times 8}{2}$}}
\psfrag{val2}{\scriptsize{$\frac{4 \times 8}{2}$}}

\begin{figure}[h]
\hspace{-3mm} 
\begin{center}
\includegraphics[]{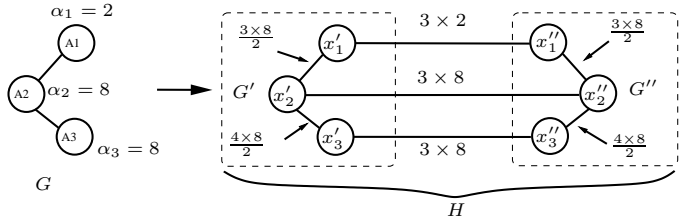}

\end{center}
\caption{Example of the transformation}
\label{fig:ExGcc2}
\end{figure}

One can show that there is a (weighted) perfect matching on $H$, which cover all the vertices of $H$. In fact the construction implies that for any perfect matching $W$ of cost $C$ on $H$,  one can provide a valid schedule of processing time $C$ for the scheduling problem : an edge $e \in W$ with $e={x',x''}, x'\in G' \wedge x''\in G''$ implies that task $x$ is scheduled alone, while an edge $e \in W$ with $e={x',y'}, x', y'\in G'$ implies that tasks $x$ and $y$ are packed together in the resulting schedule - and the edge $e={x'',y''}, x'', y''\in G''$ belong also to the matching -. 

For a minimum weight perfect matching $C$, we can associate a schedule of minimum processing time equal to $C$ and vice versa. The detailed proof of the relationship between a solution to our problem with $G$ and a solution of a minimum weight perfect matching in $H$ is presented in \cite{sdgk11journal}.

In the review of the literature, the Edmonds algorithm determines a minimum weight perfect matching in $O(n^3)$ \cite{Edmonds}. So the optimization problem with $G$ is polynomial, and if one adds the execution of the blocks created by removed vertices, this leads to the polynomiality of the problem $1| \alpha(t_i), G=chain | C_{max}$.  
\end{pf}

\subsection{Star graph}
We focus on the case with a star graph, \textit{i.e}. a graph with a central node $\beta$. In such a context, we show that the complexity depends on the number of outgoing arcs from $\beta$. The following results also imply that the studied problem can be  $\mathcal{NP}$-hard even on acyclic low-diameter graphs, when the degree of $G$ is unbounded. 

\begin{thm}
\label{corostar}
   The problem  $1|\alpha(t_i), G=star | C_{max}$ is polynomial if the central node admits at least one outcoming arc.
\end{thm}

\begin{pf}

Let $S$ be the set of satellite nodes. According to the Remark \ref{rem:IS}, $seq(S)$ is a lower bound for the cost of an optimal solution. This bound is achieved if we can execute the central node in a satellite node.

\end{pf}
 
\begin{thm}
\label{startheo}
   The problem 
  $1|\alpha(t_i), G=star | C_{max}$ is $\mathcal{NP}$-hard if the central node admits only incoming arcs.
\end{thm}

\begin{pf}
We propose a reduction from the \subsetsum (\ssp) problem  (see  Section \ref{prerequisites}). From an instance of  \ssp ~composed by a set $\mathcal{S}$ of $n$ positive values and $v \in \bbbn$ (with $v \geq x, \forall x\in \mathcal{S}$), we construct an instance of $1| \alpha(t_i),G=star|C_{max}=\sum_{t\in V}\alpha(t)+ 2\alpha(\beta)$ in the following way:
\begin{enumerate}
\item For each value $i \in \mathcal{S}$ we introduce a coupled-task $t$ with $\alpha(t)= i$. Let $V$ be the set of these tasks.
\item We add a task $\beta$ with $\alpha(\beta)=a_{\beta}=l_{\beta}=b_{\beta} = 3\times v$. 
\item We define  a compatibility constraint between each task $t \in V$ and $\beta$.
\end{enumerate}

Clearly the compatibility graph $G$ is a star with $\beta$ as the central node, and the transformation is computed in polynomial time.

We will prove that there exists a positive solution for the \subsetsum (\ssp) ~problem iff there exists a feasible solution for the scheduling problem with a length $\sum_{t\in V}\alpha(t)+ 2\alpha(\beta)$.

It is easy to see that $1| \alpha(t_i), G=star|C_{max}=\sum_{t\in V}\alpha(t)+ 2\alpha(\beta) \in \mathcal{NP}$.

Let $W$ be the set of the nodes executed in the central node for a scheduling. The cost of this scheduling in clearly $seq(\mathcal{T})-seq(W)$. Therefore, the problem of finding a scheduling of cost $seq(\mathcal{T})-\alpha(\beta)$ is clearly equivalent to an instance of the subset sum with $v=\alpha(\beta)$ and $S$ the set of the processing time of the satellite tasks. 

This concludes the Proof of Theorem \ref{startheo}.
\end{pf}

\section{On the boundary between polynomial-time algorithm  and $\mathcal{NP}$-Completeness on 1-stage bipartite graphs}

Preliminary results of Section~\ref{complexitesection} show that the problem is $\mathcal{NP}-$hard on acyclic low-diameter instances when the degree is unbounded. They suggest that the complexity of the problem may be linked to the maximum degree of the graph.

This section is devoted to the $\mathcal{NP}-$completeness of several scheduling problems in presence of a $1$-stage bipartite compatibility graph, according to the maximum degree of vertices and some structural parameters like the number of different values of coupled-tasks.

We will sharp the line of demarcation between the polynomially solvable cases and the $\mathcal{NP}$-hardness ones according to several topologies.
We focus our analysis when $G$ is a $1$-stage bipartite graph. We prove that the problem is solvable within a $O(n^3)$ polynomial algorithm if $\Delta_G=2$ (Theorem \ref{polybipartite}), but becomes \nphard when $\Delta_G= 3$ (Theorem \ref{bipartitheo}). 

We start by designing a polynomial-time algorithm for the scheduling problem in which the maximum  degree of incoming arcs on $Y$-tasks is at most two. 

\begin{thm}
  \label{polybipartite}
  The problem of deciding whether an instance of
  $1|\alpha(t_i)$, \\   $G =1-stage bipartite, d_{G}(Y) \leq 2 | C_{max}$ is polynomial. In fact, the previous result may be extended to a graph $G$ (not necessarily bipartite) such that $\forall x, d^-(X) \leq 2$ with $3(\alpha(x_1)+\alpha(x_2))>\alpha(x)$, where $x_1$ and $x_2$ are the $2$ neighbors of $x$.
\end{thm}

\begin{pf}
Let $G=(X \cup Y,E)$ be a $1$-stage~bipartite compatibility graph (arcs oriented from $X$ to $Y$ only, implying that only $X$-tasks can be executed in the idle time of and $Y$-task). $Y$-tasks will always be scheduled sequentially as $Y$ is an independent set of G (cf. Remark~\ref{rem:IS}). 
The aim is to fill their idle time with a maximum of $X$-tasks, while the remained tasks will be executed after the $Y$-tasks. We just have to minimize the length of the remained tasks.
It is easy to see that all $Y$-tasks with incoming degree equal to one are executed sequentially with their only $X$-task in their idle time.
The following algorithm is focused on the case $\Delta_{G} = 2$. 
It is defined in two steps:
  \begin{enumerate}
  \item For each task $y\in Y$ such that $3 \times \alpha(x_1) +3 \times \alpha(x_2) \leq \alpha(y)$ where $x_1$ and $x_2$ are the only two neighbors of $Y$, we add $y$ to the schedule and execute $x_1$ and $x_2$ sequentially during the idle time of $y$. Then we remove $y$, $x_1$ and $x_2$ from the instance.
  
\item Each remaining task $y \in Y$ admits at most two incoming arcs ($x_1,y)$ and/or $(x_2,y)$.  We add a weight $\alpha(x)$  to the arc $(x,y)$ for each $x\in N_G(y)$, then we perform a maximum weight matching on $G$ in order to minimize the length of the remained tasks of $X$. Thus, the matched 
coupled-tasks  are executed, and these tasks are removed from $G$. 

\item Then, remaining tasks are processed sequentially after the other tasks.
\end{enumerate}
The complexity of this algorithm is $O(n^3)$ using the Hungarian method \cite{hungarian}. For the extension, it is sufficient to use a maximum weight perfect matching \cite{Edmonds}.
\end{pf}

\begin{thm}
\label{bipartitheo}
  The problem of deciding whether an instance of
   $1|\alpha(t_i), G =$ \\ $1-stage~bipartite,d_{G}(X)=2, d_{G}(Y)\in \{2,3\} | C_{max}$ has a schedule of length at most $54n$ is $\mathcal{NP}$-complete where $n$ is the number of tasks.
\end{thm}

\begin{pf}
It is easy to see that our problem is in
${\mathcal{NP}}$.
Our proof is based on a reduction  from \SAToneinthree: given 
a set $\mathcal{V}$ of $n$ boolean variables with $n\mod 3\equiv 0$, a set of $n$ clauses of cardinality two and $n/3$ clauses of cardinality three, we construct an instance ${\mathcal{\pi}}$ of the problem
$1|\alpha(t_i), G=$ $1-stage~bipartite, d_{G}(X)=2, d_{G}(Y)\in \{2,3\} | C_{max} =54n$ in following way:

\begin{enumerate}
\item For all $x \in {\mathcal{V}}$, we introduce four variable-tasks:
  $x, ~ x',~ \bar{x}$ and $\bar{x}'$ with $(a_i,l_i,b_i)=(1,1,1), \forall i
  \in \{x,x',\bar{x}, \bar{x}'\}$. This  variable-tasks set is noted $\mathcal{VT}$.
\item For all  $x \in {\mathcal{V}}$,
  we introduce three literal-tasks $\mathcal{L}_x, C^x$ and
  $\bar{C}^x$ with $\mathcal{L}_x=(2,2,2); C^x=\bar{C}^x=(6,6,6)$. The set of literal-tasks is
  denoted $\mathcal{LT}$.
\item For all clauses with a length of three, we introduce two clause-tasks $C^i$ and $\bar{C}^i$ with
$C^i=(3,3,3)$ and $\bar{C}^i=(6,6,6)$. 
\item For all clauses with a  length of two, we  introduce one clause-task $C^i$ with $C^i=(3,3,3)$.
The set of clause-tasks is
  denoted $\mathcal{CT}$.
\item The following arcs model the compatibility constraint:
  \begin{enumerate}
  \item For all boolean variables  $x \in {\mathcal{V}}$, we add the arcs  $(\mathcal{L}_x , C^x)$
  and $(\mathcal{L}_x, \bar{C}^x)$
\item For all clauses with a length of three denoted $C_i=(y \vee z
  \vee t)$, we add the arcs  
$(y, C^i)$, $(z , C^i)$, $(t,C^i)$ and $(\bar{y}' ,\bar{C}^i)$, $(\bar{z}', \bar{C}^i)$, $(\bar{t}', \bar{C}^i )$.
\item For all clauses with a  length of two denoted $C_i=(x \vee \bar{y})$, we 
add  the arcs $(x', C^i)$ and $(\bar{y},  C^i)$.
\item Finally, we add the arcs $(x, C^x)$, $(x', C^x)$ and
  $(\bar{x}, \bar{C}^x)$, $(\bar{x}', \bar{C}^x)$.
  \end{enumerate}
\end{enumerate}

\psfrag{xx}{$x$}\psfrag{yy}{$y$}\psfrag{zz}{$z$}
\psfrag{primex}{$x'$}
\psfrag{barx}{$\bar{x}$}
\psfrag{barprimex}{$\bar{x}'$}\psfrag{barprimey}{$\bar{y}'$}\psfrag{barprimez}{$\bar{z}'$}
\psfrag{Lx}{$\mathcal{L}_x$}
\psfrag{Cx}{$C^x$}
\psfrag{barCx}{$\bar{C}^x$}
\psfrag{clause3}{$(x,y,z)$}
\psfrag{barclause3}{$\overline{(x,y,z)}$}
\psfrag{CC}{$C$}
\psfrag{CCprime}{$C'$}
\psfrag{texte}{$C \neq C'$ two clause-tasks of length two}
\psfrag{blabla}{$(a_i,l_i,b_i),=(1,1,1), \forall i
  \in \{x,x',\bar{x}, \bar{x}'\}, \forall x \in {\mathcal{V}}$}
\psfrag{blabla1}{$\mathcal{L}_x=(2,2,2); C^x=\bar{C}^x=(6,6,6), x \in
  {\mathcal{V}}$}
\psfrag{blabla2}{$C^{(x,y,z)}=(3,3,3)$}
\psfrag{blabla2bis}{$C^{\overline{(x,y,z)}}=(6,6,6)$}
\psfrag{blabla3}{$C^C=C^{C'}=(3,3,3)$}
\psfrag{casa}{Case $a)$}
\psfrag{casb}{Case $b)$}
\psfrag{xrecoit}{$x$ is true and $\bar{x}$ is false}
\psfrag{xrecoit1}{$x$ is false and $\bar{x}$ is true}
\afig{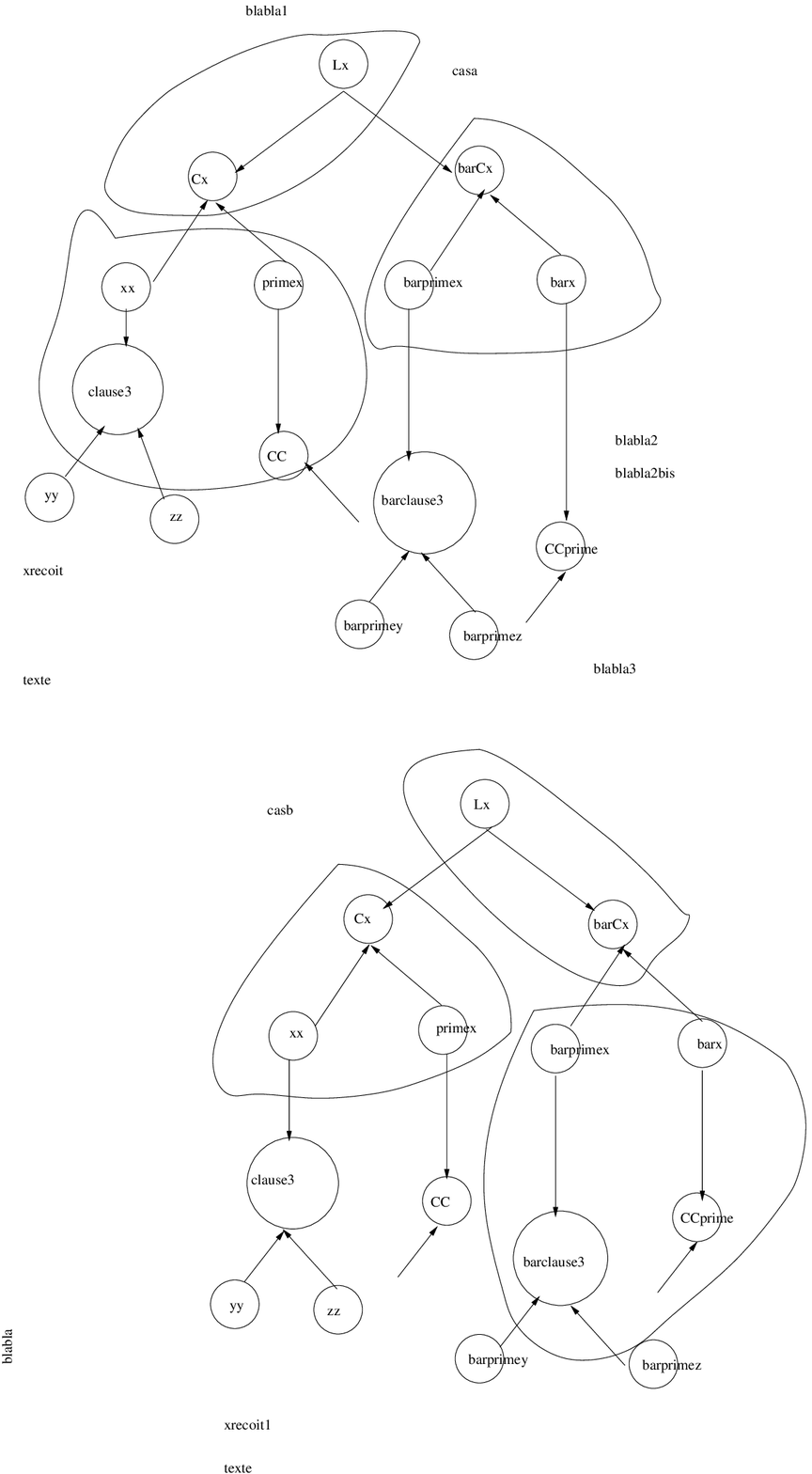}{A partial compatibility graph for  the
  ${\mathcal{NP}}$-completeness of the scheduling problem
  $1|\alpha(t_i),G =$1$-stage~bipartite, d_{G}(X)=2, d_{G}(Y)\in \{2,3\}| C_{max}=54n$. A truth assignment and partial schedule.}

\noindent This transformation can be clearly computed in polynomial time and an illustration is depicted in Figure \ref{fig:construction2}. The proposed compatibility graph is a $1$-stage bipartite 
and $d_{G}(x) \leq 3, \forall x \in \mathcal{VT} \cup \mathcal{LT}\cup  \mathcal{CT}$.

In follows, we say that a task $x$ is merged to a task $y$, if there
exists a compatibility constraint from $x$ to $y$;  {\it  i.e.} the coupled-task $x$ may be executed during the idle time of coupled-task $y$. 
  
\begin{bulletitemize}
\item  Let us first assume that there is a schedule with length of  $54n$ at most.
We prove that there is a truth assignment $I: {\mathcal{V}} \rightarrow \{0,1\}$  
such that each clause in ${\mathcal{C}}$ has exactly one true literal (\textit{i.e.} one literal equal to $1$).

We make several essential remarks:

\begin{enumerate}
\item The length of the schedule is given by an execution time of the
  coupled-tasks admitting only incoming arcs, and the value is 
 $54n=3\alpha_{\mathcal{CT}}|\mathcal{CT}| + \alpha_{\mathcal{LT}}(|\mathcal{LT}|
  - |\{\mathcal{L}_x, x \in \mathcal{V}\}|) = 9\vert \{C^i \in \mathcal{CT}$ of length $2$ and $3 \}\vert +
  18\vert \{ \bar{C}^i \in \mathcal{CT} \}\vert + 18 \vert\{ C^x$ and  $\bar{C}^x \in \mathcal{LT}\} \vert
= 9\times \frac{4n}{3} + 18\times \frac{n}{3} + 18\times 2n$.

   Thus, all tasks from $\mathcal{VT} \cup \{\mathcal{L}_x, x \in \mathcal{V}\}$ must be
  merged with tasks from $\mathcal{CT} \cup (\mathcal{LT} - \{\mathcal{L}_x, x \in \mathcal{V}\})$.
\item By the construction, at most three tasks can be merged together.
\item $\mathcal{L}_x$ is merged with $C^x$ or $ \bar{C}^x$.
\item The allocation of coupled-tasks from $\mathcal{CT} \cup (\mathcal{LT} 
- \{\mathcal{L}_x, x \in \mathcal{V}\})$ leads to  $18n$ idle
  time. The length of the variable-tasks $\mathcal{VT}$ and $\mathcal{L}_x$ equals
   $18n$ (in these coupled-tasks there are $6n$ idle times).

\item If the variable-tasks $x$ and  $x'$ are not merged simultaneously
  with $C^x$, {\it  i.e.} only one of these tasks is merged with $C^x$,
  then by with the previous discussion, it is necessary to merge a literal-task
  $\mathcal{L}_y$, with $x \neq y$ one variable-task ($\bar{y}$ or $\bar{y}'$) with
  $C^y$ or $\bar{C}^y$. It is impossible by size of coupled-tasks. In the same way, the variable-tasks $\bar{x}$ et $\bar{x}'$ are merged simultaneously with $\bar{C}^x$ if they have to be into it. 
\item Hence, first  $x$ and $x'$ are merged with $C^x$ or with a clause-task where the variable $x$ occurs. Second, $\bar{x}$ and $\bar{x}'$ are merged with $\bar{C}^x$ or a clause-task.
\end{enumerate}

So, we affect the value "true" to the variable $l$ iff the
variable-task $l$ is merged with clause-task(s) corresponding to 
the clause where the variable $l$
occurs. It is obvious to see that in the clause of length three and two we have one and only one literal equal to "true''.

\item Conversely, we suppose that there is a truth assignment $I: {\mathcal{V}} \rightarrow \{0,1\}$,  such that each clause in ${\mathcal{C}}$ has exactly one true literal.

\begin{itemize}
\item If the variable $x=true$ then we merged the vertices $\mathcal{L}_x$ with $C^x$; $x$ with the clause-task $C^i$ corresponding to the clause of length three which $x$ occurs; $x'$ with  the clause-task $C^i$ corresponding to the clause of length two which $x$ occurs; and $\bar{x}, \bar{x}'$ with $\bar{C}^x$.
\item  If the variable $x=false$ then we merged the vertices $\mathcal{L}_x$ with $\bar{C}^x$; $\bar{x}$ with the clause-task corresponding to the clause of length
two which $\bar{x}$ occurs;
$\bar{x}'$ with  the clause-task $\bar{C}^i$ corresponding to the
  clause ($C$) of length
  three which $x$ occurs; and $x, x'$ with $C^x$.
\end{itemize}
The  merged-tasks are given in Figure \ref{fig:construction2}.
 For a
feasible schedule, it is sufficient to merge vertices which are in the same partition.
Thus, the length of the schedule is at most $54n$.

  \end{bulletitemize}

\end{pf}

\begin{thm}
\label{bipartitheobis}
  The problem of deciding whether an instance of
   $1|\alpha(t_i), G =$ \\ $1-stage~bipartite,d_{G}(X) \in \{1,2\}, d_{G}(Y)\in \{3,4\} | C_{max}$ has a schedule of length at most $54n$ is ${\mathcal{NP}}$-complete, where $n$ is the number of tasks.
\end{thm}

\begin{pf}
We use a similar proof as given for the Theorem \ref{bipartitheo}. It is sufficient to add for each clause $C$ with a  length of two (resp.  $C'$ of length three) a dummy coupled-task $\mathcal{D}_C$ (resp. $\mathcal{D}_C'$) with $\mathcal{D}_C=(1,1,1)=\mathcal{D}_C'$, and the value of the   clause-task $C$ (resp. $C'$) is now $C=C'=(6,6,6)$. In other words, we add these two compatibility constraints:

\begin{itemize}
\item $\mathcal{D}_C \rightarrow C$, for each clause $C$ of length two,
\item $\mathcal{D}_C' \rightarrow C'$, for each clause $C'$ of length three.
\end{itemize}

There is a schedule with length of  $54n$ at most iff there exists a 
truth assignment $I: {\mathcal{V}} \rightarrow \{0,1\}$  
such that each clause in ${\mathcal{C}}$ 
has exactly one true literal (\textit{i.e.} one literal equal to $1$).
\end{pf}
    
\begin{cor}
 \label{extension}
The problem of deciding whether an instance of
  $1| \alpha(t_i) \neq \alpha(t_j), \forall i \neq j, \Delta_G =3, G= 1-stage ~bipartite  |C_{max}$ has a schedule of length at most $54n$ is ${\mathcal{NP}}$-complete, where $n$ is the number of tasks.
\end{cor}

\begin{pf}
The proof of Theorem \ref{bipartitheo} can be adapted by using the
classical scaling arguments assigning $\alpha(x)+ \epsilon$ to each task.
\end{pf}

\section{Polynomial-time approximation algorithms}
\label{approxSection}

This section is devoted to design some efficient polynomial-time approximation algorithms for several topologies and mainly for bipartite graphs.
In \cite{sgk10}, authors proposed a simple algorithm, which consists in scheduling all the tasks consecutively, with a performance ratio bounded by $3/2$ for a general compatibility graph. 
The challenge for the remaining section, is to propose some efficient algorithms with  a ratio strictly lower than $3/2$.
We propose a $\mathcal{FPTAS}$ for the star graph whereas some $\mathcal{APX}$-algorithms are developed in the remaining section according to the characteristics  of the $1$-stage bipartite graph. At last, we extend the result is extended to the $2$-stage bipartite graph.

\subsection{Star graph}

\begin{thm}
\label{startheoapprox}
   The problem   $1| \alpha(t_i), G=star | C_{max}$ admits a $\mathcal{FPTAS}$.
\end{thm}

\begin{pf}
The central node admits only incoming arcs (the case of the central node admits at least one  outcoming arc is given by Corollary \ref{corostar}). Therefore,  we may use the solution given by the \subsetsum (\ssp)  (see \cite{ibarra} and \cite{kellerer}). Indeed, based on the reduction used in the proof of Theorem \ref{startheo} and the optimization version  of \ssp: the aim is to find $W^*$ (an optimal set of tasks executed during the idle time of the central node) which maximizes $seq(W^*)$ such that $seq(W^*) \leq \alpha(\beta)$.

Let us suppose that $\frac{seq(W)}{seq(W^*)} \geq 1 -\epsilon$, where $W$ designates the value of the approximation solution for \subsetsum\!\!. 

Note that $\alpha{(\beta)} \geq seq(W^*)$ and $seq(\mathcal{T}) \geq 3 \alpha(\beta)$ lead to $seq(\mathcal{T}) \geq 3seq(W^*)$.

\begin{eqnarray*}
\frac{seq(\mathcal{T})-seq(W)}{seq(\mathcal{T})-seq(W^*)} &=& 1+ \frac{seq({W^*})-seq(W)}{seq(\mathcal{T})-seq(W^*)}\\
& \leq & 1+ \frac{seq({W^*})-seq(W)}{2seq(W^*)}\\
&\leq & 1+\frac{1-\frac{seq(W)}{seq(W^*)}}{2}
\leq  1+\frac{1-(1-\epsilon)}{2}=1+\epsilon/2\\
\end{eqnarray*}
Therefore the existence of a $\mathcal{FPTAS}$ for the \subsetsum involves a $\mathcal{FPTAS}$ for our scheduling problem.

\end{pf}







\subsection{1$-$stage~bipartite graph}
Scheduling coupled-tasks during the idle time of others can be related to packing problems, especially when the compatibility graph $G$ is a bipartite graph. In the following, we propose several approximations when  $G$ is a 1$-$stage~bipartite graph.

\begin{lem}
Let $\Pi$ be a problem with  $\Pi \in \{\MKAR, \MSSDC,\MSS\}$ such that $\Pi$ admits a $\rho$-approximation, then the following problems 
\begin{enumerate}
\item  $1| \alpha(t_i), G=1-stage~bipartite | C_{max}$,
\item $1| \alpha(t_i), G=complete ~ 1-stage~ bipartite | C_{max}$,
\item  $1 | \alpha(t_i), G=complete~ 1-stage~ bipartite | C_{max}$, 
where $G=(X \cup Y,E)$ and all the tasks from $Y$ have the same stretch factor $\alpha(Y)$,
\end{enumerate}
 posses a $\rho'$-approximables within a factor $\rho'=1 + \frac{(1-\rho)}{3}$ using an approximability reduction from $\MKAR, \MSSDC$ and $\MSS$ respectively.
\label{lemmaall}
\end{lem}

\begin{pf}
\begin{enumerate}
\item Consider now an instance of $1 | \alpha(t_i), G =1-stage~bipartite | C_{max}$ with a graph $G=(X\cup Y,E)$ (for any arc $e=(x,y)\in E$, we have $x\in X$ and $y\in Y$) and a stretch factor function $\alpha: X\cup Y\rightarrow \bbbn$.

In such an instance, any valid schedule consists in finding for each task $y\in Y$ a subset of compatible tasks $X_y\subseteq X$ to pack into $y \in Y$, each task of $x$ being packed at most once.  Let $X_p= \cup_{y\in Y} X_y$ be the union of $X$-tasks packed into a task from $Y$. 
Let  $X_{\bar{p}}$ the set of remaining tasks, with $X_{\bar{p}}  = X \setminus X_p$.  Obviously, we have:
\begin{equation}
seq(X_p) + seq( X_{\bar{p}})  = seq(X)
\label{eq_01}
\end{equation}

As $Y$ is an independent set in $G$, $Y$-tasks have to be scheduled sequentially in any (optimal) solution.  The length of any schedule $S$ is then the time required to execute entirely the $Y$-tasks plus the one required to schedule sequentially the tasks from $X_{\bar{p}}$. Formally: 
\begin{equation}
C_{max}(S) = seq(Y) + seq(X_{\bar{p}})
\label{eq_02}
\end{equation}

From Equations (\ref{eq_01}) and (\ref{eq_02}) we have:
\begin{equation}
C_{max}(S) = seq(Y) + seq(X) - seq(X_p).
\label{eq_03} 
\end{equation}

We use here a transformation into a \MKAR ~problem:  each task $x$ from $X$ is an item having a weight $3\alpha(x)$, each task $y$ from $Y$ is a bin with a capacity $\alpha(y)$, and each item $x$ can be packed into $y$ if and only if the edge $(x,y)$ belongs to the bipartite graph.

Using algorithms and results from the literature, one can obtain an assignment of some items into bins.
We denote by $X_p$ the set of these packed items. 
The cost of the solution for the \MKAR ~problem is $seq(X_p)$. If \MKAR ~is approximable to a factor $\rho$, then we have:
\begin{equation}
 seq(X_p)\geq \rho \times seq(X^*_p),
\label{eq_04}
\end{equation} 
 where  $X^*_p$ is the set of packable items with the maximum profit.
Combining Equations (\ref{eq_03}) and (\ref{eq_04}),  we obtain  a solution for $1 |\alpha(t_i), G=1-stage\ bipartite | C_{max}$  with a length:
\begin{equation}
C_{max}(S) \leq seq(Y) + seq(X) - \rho \times seq(X^*_p)
\label{eq_05} 
\end{equation} 

As $X$ and $Y$ are two fixed sets, an optimal solution $S^*$  with minimal length $C_{max}(S^*)$ is obtained when $seq(X_p)$ is maximum, {\it  i.e.} when $X_p=X^*_p$.  The length of any optimal solution is equal to: 
\begin{equation}
C_{max}(S^*) = seq(Y) + seq(X) - seq(X^*_p)
\label{eq_06} 
\end{equation}

Using Equations (\ref{eq_05}) and (\ref{eq_06}), the ratio obtained between our solution $S$ and the optimal one $S^*$ is:
\begin{equation}
\frac{C_{max}(S)}{C_{max}(S^*)} \leq  \frac{seq(Y) + seq(X) - \rho \times seq(X^*_p)}{seq(Y) + seq(X) - seq(X^*_p)} \leq  1 + \frac{(1-\rho)\times seq(X^*_p)}{seq(Y) + seq(X) - seq(X^*_p)}
\label{eq_07} 
\end{equation}

By definition,  $X^*_p\subseteq X$. Moreover, as the processing time of  $X^*_p$ cannot excess the idle time of tasks from $Y$,  we obtain:
\begin{equation}
seq(X^*_p) \leq \frac{1}{3}seq(Y) 
\label{eq_08} 
\end{equation}

Combined to Equation  (\ref{eq_07}), we obtain the desired upper bound:

\begin{equation}
\rho'=\frac{C_{max}(S)}{C_{max}(S^*)}  \leq  1 + \frac{(1-\rho)}{3}.
\label{eq_09} 
\end{equation}\\

\item For the problem $1 | \alpha(t_i), G=complete~ $1$-stage~ bipartite | C_{max}$, the proof is similar to the previous one. We remind that \MSSDC ~is a special case of \MKAR ~in which each item can be packed in any bin. \\

\item For the problem $1 | \alpha(t_i), G=complete~ 1-stage~ bipartite | C_{max}$  where $G=(X \cup Y, E)$ and all the $Y$-tasks have the same stretch factor $\alpha(Y)$, the proof is similar to the previous one since \MSSDC ~is a generalization of \MSS. 

\end{enumerate}
\end{pf}

\begin{thm}
\label{alltheorems}
The following problems admit a polynomial-time approximation algorithm:
\begin{enumerate}
\item The problem $1| \alpha(t_i), G=1-stage~ bipartite | C_{max}$  is approximable to a factor $\frac{7}{6}$.
\item The problem $1| \alpha(t_i), G=complete~ 1-stage~ bipartite | C_{max}$  admits a $\mathcal{PTAS}$.
\item The problem $1| \alpha(t_i), G=complete~ 1-stage~ bipartite | C_{max}$ where $G=(X \cup Y, E)$  and all the $Y$-tasks have the same stretch factor $\alpha(Y)$:
\begin{enumerate}
\item is approximable to a factor $\frac{13}{12}$.
\item admits a $\mathcal{PTAS}$.
\end{enumerate}

\end{enumerate}

\end{thm}


\begin{pf}

\begin{enumerate}
\item Authors from \cite{DKKSR00} proposed a $\rho=\frac{1}{2}-$approximation algorithm for \MKAR. Reusing this result  with Lemma~\ref{lemmaall}, we obtain a $\frac{7}{6}-$approximation for $1| \alpha(t_i), G=1-stage~ bipartite | C_{max}$.

\item We know that \MSSDC ~admits a $\mathcal{PTAS}$  \cite{CKP002}, {\it  i.e.} $\rho=1-\epsilon$. Using this algorithm to compute such a $\mathcal{PTAS}$, with  Lemma \ref{lemmaall} we obtain an approximation ratio of $1 + \frac{\epsilon}{3} $ for this problem.
\item 
In this case we have two different results:
\begin{enumerate}
\item Authors from \cite{CKP03} proposed a $\rho=\frac{3}{4}-$approximation algorithm for \MSS. Reusing this result  with Lemma~\ref{lemmaall}, we obtain a $\frac{13}{12}-$approximation for $1 |\alpha(t_i), G=complete~ bipartite | C_{max}$. 
\item They also  proved that \MSS~ admits a $\mathcal{PTAS}$ \cite{CKP00} , {\it  i.e.} $\rho=1-\epsilon$. Using the algorithm to compute such a $\mathcal{PTAS}$, with Lemma \ref{lemmaall} we obtain an approximation ratio of $ 1 + \frac{\epsilon}{3}$ for $1|\alpha(t_i), G=complete~ 1-stage ~bipartite | C_{max}$ when $Y$-tasks have the same stretch factor.
\end{enumerate}

\end{enumerate}

\end{pf}

\subsection{2$-$stage~bipartite graph}
\label{section2bipartite}

In the following, we extend the previous result for $2$-stage bipartite graphs. 
\begin{thm}
  \label{th:2stage:apx13/9}
  The problem $1| \alpha(t_i), G=2-stage~ bipartite | C_{max}$  is approximable to a factor $\frac{13}{9}$.
\end{thm}

\begin{pf}
The main idea of the algorithm is divided into three steps:
\begin{enumerate}
\item First we compute a part of the solution with a $\frac{7}{6}$-approximation on $G_0$ thanks to Theorem \ref{alltheorems}, where $G_0=G[V_0 \cup V_1]$ is the $1$st stage of $G$.
\item Then we compute a second part of the solution with a $\frac{7}{6}$-approximation on $G_1$ thanks to Theorem \ref{alltheorems}, where $G_1\!=\!G[V_1 \!\cup\! V_2]$ is  the $2$nd stage of $G$.
\item To finish we merge these two parts and we resolve potential conflicts between them, {\it  i.e.} by giving a preference to tasks packed in $G_1$.
Computing the cost of this solution gives us an approximation ratio of $\frac{13}{9}$.
\end{enumerate}

Reusing the notation introduced for $k$-stage bipartite graphs (see Section \ref{notations}), we consider an instance of  $1| \alpha(t_i), G=2-stage~ bipartite | C_{max}$ with $G=(V_0 \cup V_1 \cup V_2, E_1 \cup E_2)$, where  each arc in $E_i$ has its extremities in $V_{i-1}$ and $V_{i}$, for $i \in \{1,2\}$.

\begin{def}
\label{def}
$\forall i=\{0,1\}$ we denote\footnotemark[1] by ${V_i}_p$ (resp. ${V_i}_a$) the set of tasks merged (resp. remaining) in any task from $y \in V_{i+1}$ in a solution $S$. 
Moreover, $\forall i=\{1,2\}$ let ${V_i}_b$ be the set of tasks scheduled with some tasks from $V_{i-1}$ merged into it. This notation is extended to an optimal solution  $S^*$ by adding a star in the involved variables.
\end{def}

\footnotetext[1]{Notations: p for packed, a for alone, and b for box}   

Considering the specificities of the instance, in any (optimal) solution we propose some essential remarks:
\begin{enumerate}
\item Tasks from $V_0$ are source nodes in $G$, and they can be either scheduled alone, or merged only into some tasks from $V_1$ only. Given any solution $S$ to the problem on $G$, $\{{V_0}_p, {V_0}_a\}$ is a partition of $V_0$.

\item Tasks from $V_1$ can be either scheduled alone, merged into some tasks from $V_2$,  or scheduled with some tasks from $V_0$ merged into it. Given any solution $S$ to the problem on $G$, $\{{V_1}_p, {V_1}_a, {V_1}_b\}$ is a partition of $V_1$.

\item Tasks from $V_2$ form an independent set in $G$, and have to be scheduled sequentially in any schedule (cf. Remark~\ref{rem:IS}), either alone or with some tasks from $V_1$ merged into it. Given any solution $S$ to the problem on $G$,  $\{{V_2}_a, {V_2}_b\}$ is a partition of ${V_2}$.
\end{enumerate} 
Any solution would consist first to schedule each task with at least one task merged into it, then to schedule the remaining tasks (alone) consecutively. 
Given an optimal solution $S^*$, the length of $S^*$ is given by the following equation: 
\begin{equation}
\label{eq:4}
S^*=  seq({V_1}^*_b) +  seq({V_2}_b) + seq({V_0}^*_a) + seq({V_1}^*_a)  + seq({V_2}^*_a) 
\end{equation}
or, more simply 
\begin{equation}
S^*=  seq({V_2}) + seq({V_1}^*_b)  + seq({V_0}^*_a) + seq({V_1}^*_a) 
\label{eq:borneinfS*}
\end{equation}

Note that ${V_0}^*_p$ and ${V_1}^*_p$ are not part of the equation, as they are scheduled during the idle time of  ${V_1}^*_b$ and ${V_2}^*_b$.

The main idea of the algorithm is divided into three steps:
\begin{enumerate}
\item First we compute a part of the solution with a $\frac{7}{6}$-approximation on $G_0$ thanks to Theorem \ref{alltheorems}, where $G_0=G[V_0 \cup V_1]$ is  the $1$st stage of $G$.
\item Then we compute a second part of the solution with a $\frac{7}{6}$-approximation on $G_1$ thanks to Theorem \ref{alltheorems}, where $G_1\!=\! G[V_1 \!\cup\! V_2]$ is  the $2$nd stage of $G$.
\item To finish we merge these two parts and we solve potential conflicts between them.
\end{enumerate}

Let consider an instance restricted to the graph $G_0$. Note that  $G_0$ is a $1$-stage bipartite graph. 
Let $S^*[G_0]$ be an optimal solution on $G_0$. Let us denote by ${V_0}^*_p[G_0]$ the set of tasks from $V_0$ packed into tasks from $V_1$ in $S^*[G_0]$, and by ${V_0}^*_a[G_0]$ the set of remaining tasks.

Obviously, we have:
\begin{equation}
S^*[G_0] = seq(V_1) + {V_0}^*_a[G_0]
\end{equation}

Given any solution $S[G_0]$, let ${V_1}_b[G_0]$ be the set of tasks from $V_1$ with at least one task from $V_0$ merged into them, and ${V_1}_a[G_0]$ be the remaining tasks. Let us denote by ${V_0}_p[G_0]$ the set of tasks from $V_0$ merged into $V_1$, and by ${V_0}_a[G_0]$ the set of remaining tasks.  Using Theorem \ref{alltheorems}, Lemma \ref{lemmaall}, and the demonstration presented in the proof from \cite{DKKSR00}, we compute a solution $S[G_0]$  such that:
\begin{equation}
seq({V_0}_p[G_0]) \geq \frac{1}{2} seq({V_0}^*_p[G_0] )
\label{eq_0001}
\end{equation}
Note that we have:
\begin{equation}
seq({V_0}_p[G_0] ) + seq({V_0}_a[G_0] ) = seq({V_0}^*_p[G_0] )+ seq({V_0}^*_a[G_0])  = seq(V_0)
\label{eq_0002}
\end{equation}
Combining Equations (\ref{eq_0001}) and (\ref{eq_0002}), we obtain:
\begin{equation}
seq({V_0}_a[G_0])  \leq seq({V_0}^*_a[G_0])  + \frac{1}{2} seq({V_0}^*_p[G_0])  \leq seq({V_0}^*_a)  + \frac{1}{2} seq({V_0}^*_p[G_0])  
\label{eqborne1}
\end{equation}
as we know  by definition that $seq({V_0}^*_a[G_1]) \leq seq({V_0}^*_a)$.\\

We use a similar reasoning on  an instance restricted to the graph $G_1$. 
Let $S^*[G_1]$ be an optimal solution on $G_1$. Let us denote by ${V_1}^*_p[G_1]$ the set of tasks from $V_1$ packed into tasks from $V_2$ in $S^*[G_1]$, and by ${V_1}^*_a[G_1]$ the set of remaining tasks. 
Given any solution $S[G_1]$, let ${V_2}_b[G_1]$ be the set of tasks from $V_2$ with at least one task from $V_1$ merged into them, and ${V_1}_a[G_1]$ be the remaining tasks.  One can compute a solution $S[G_1]$ based on a set of tasks ${V_1}_p[G_1]$ packed in $V_2$ such that:
\begin{equation}
seq({V_1}_p[G_1]) \geq \frac{1}{2} seq({V_1}^*_p[G_1] )
\label{eq_0004}
\end{equation}
and 
\begin{equation}
seq({V_1}_a[G_1])  \leq seq({V_1}^*_a[G_1])  + 1/2 seq({V_1}^*_p[G_1])  \leq seq({V_1}^*_a)  + 1/2 seq({V_1}^*_p[G_1])
\label{eqborne1'}
\end{equation}
as we know  by definition that $seq({V_1}^*_a[G_1]) \leq seq({V_1}^*_a)$.

We design the feasible solution $S$ for $G$ as follows:
\begin{itemize}
\item First we compute a solution $S[G_1]$ on $G_1$, then we add  to $S$ each task from 
$V_2$ and the tasks from $V_1$ merged into them (i.e. ${V_1}_p[G_1]$) in $S[G_1]$.
\item Second we compute a solution $S[G_0]$ on $G_0$, then we add to $S$ each task $v$ from ${V_1}_b[G_0] \setminus {V_1}_p[G_1]$ and the tasks from $V_0$ merged into them.
\item Third the tasks  ${V_1}_a[G_1] \setminus {V_1}_b[G_0] $  and  ${V_0}_a[G_0] $  are added to $S$ and scheduled sequentially.
\item At last we denote by $V_{conflict}$ the  set of remaining tasks, {\it  i.e.} the set of tasks from $V_0$ which are merged into a task $v\in V_1$ in $S[G_0]$, thus that $v$ is merged into a task from $V_2$ in $S[G_1]$.
\end{itemize}

Observe that:
\begin{equation}
seq({V_1}_b[G_0] \setminus {V_1}_p[G_1]) +  seq({V_1}_a[G_1] \setminus {V_1}_b[G_0])  = {V_1}_a[G_1])  
\end{equation}
Thus the cost of our solution $S$ is:
 \begin{equation}
S = seq(V_2) +  seq({V_1}_a[G_1])  + seq({V_0}_a[G_0]) + seq (V_{conflict})
\label{eq:S01}
\end{equation}
 It is also clear that:
\begin{equation} 
\label{eq:conflit}
 seq(V_{conflict}) \leq \frac{1}{3} seq({V_1}_p[G_1]) \leq \frac{1}{3} seq({V_1}^*_p[G_1])
\end{equation}

Using Equations (\ref{eqborne1}), (\ref{eqborne1'}) and (\ref{eq:conflit}) in Equation (\ref{eq:S01}), we obtain:
\begin{equation}
S \leq  seq(V_2) +  seq({V_1}^*_a)  + \frac{5}{6} seq({V_1}^*_p[G_1])   +  seq({V_0}^*_a)+ \frac{1}{2} seq({V_0}^*_p[G_0]) 
\label{eq:bornesol}
\end{equation}

Using Equations (\ref{eq:borneinfS*}) and (\ref{eq:bornesol}), we obtain:
\begin{equation}
S \leq  S^* + \frac{5}{6} seq({V_1}^*_p[G_1])  + \frac{1}{2} seq({V_0}^*_p[G_0]) 
\label{eq:bornesol2}
\end{equation}

We know that $S^*\geq seq(V_2)$, and that tasks from ${V_1}^*_p[G_1]$ must be merged into tasks from $V_2$ and cannot exceed the idle time of $V_2$, implying that $seq({V_1}^*_p[G_1])) \leq \frac{1}{3} seq(V_2)$. We can write the following:
\begin{equation}
\label{eq:premMemb}
\frac{\frac{5}{6} seq({V_1}^*_p[G_1])}{S^*} \leq \frac{\frac{5}{6} \times \frac{1}{3} seq(V_2)}{seq(V_2)} \leq \frac{5}{18}
\end{equation}

We  know that tasks from ${V_0}^*_p[G_0]$ must be merged into tasks from $V_1$ and cannot exceed the idle time of $V_1$, implying that $seq({V_0}^*_p[G_0]) \leq \frac{1}{3} seq(V_1)$.
We also know that  $S^*\geq seq(V_1)$, as $V_1$ is an independent set of $G$. One can write the following:
\begin{equation}
\label{eq:secMemb}
\frac{\frac{1}{2} seq({V_0}^*_p[G_0])}{S^*} \leq \frac{ \frac{1}{2}\times \frac{1}{3} seq(V_1)}{seq(V_1)} \leq \frac{1}{6}
\end{equation}

Finally, with Equations (\ref{eq:bornesol2}), (\ref{eq:premMemb}) and (\ref{eq:secMemb})  we conclude the proof:
\begin{equation}
\frac{S}{S^*} \leq  1 + \frac{5}{18} + \frac{1}{6}  = \frac{13}{9}
\end{equation}


\end{pf}

\section{Conclusion}
In this paper, we investigate a particular coupled-tasks scheduling problem $1\vert a_i\!=\!l_i\!=\!b_i,G\vert C_{max}$ in presence of a compatibility graph with regard to the complexity and approximation. We also establish the $\mathcal{NP}$-completeness for the specific case where there is a bipartite compatibility graph. In such context, we propose a  $\frac{7}{6}$-approximation algorithm and the bound is tight. We extend the result to the $2$-stage bipartite by designing a $13/9$-approximation.


A further interesting question  concerns the study of the complexity on tree graphs with bounded degree. As we known,  no complexity result exists. Another perspective consists in extending the presented results to $k$-stage bipartite graphs. 
\enlargethispage{1cm}
\bibliographystyle{plain}
\bibliography{biblio}

\end{document}